\documentclass[aps,prx,reprint,superscriptaddress,nofootinbib,longbibliography]{revtex4-2}
\usepackage{amsmath,amssymb,graphicx,bm}
\usepackage{hyperref}
\usepackage{xcolor}
\usepackage[percent]{overpic}
\usepackage{booktabs}
\usepackage{comment}

\graphicspath{{Figures/}}

\DeclareSymbolFont{matha}{OML}{txmi}{m}{it}
\DeclareMathSymbol{\varv}{\mathord}{matha}{118}

\begin{document}

\title{Coarse-grained Shannon entropy of random walks with shrinking steps}

\author{Alexander Feigel}
\email{alexander.feigel@mail.huji.ac.il}
\affiliation{The Racah Institute of Physics, The Hebrew University of Jerusalem, Edmond J. Safra Campus, Jerusalem 9190401, Israel}

\author{Alexandre V. Morozov}
\email{morozov@physics.rutgers.edu}
\affiliation{Department of Physics and Astronomy, Rutgers, The State University of New Jersey, 136 Frelinghuysen Road, Piscataway, New Jersey 08854, USA}


\begin{abstract}
In one-dimensional diffusive processes with discrete steps characterized by geometrically decaying magnitudes,
the usual Gaussian broadening familiar from Brownian motion is replaced by
bounded probability distributions over particle positions that are characterized by multi-scale fractal structures.
In this work, we study random walks with shrinking steps (known as Bernoulli convolutions), focusing on their behavior in the vicinity of the dyadic
contraction ratio $1/2$. Our analytical and numerical results show that
the coarse-grained Shannon entropy of particle distributions induced by Bernoulli convolutions
exhibits a local maximum at the dyadic ratio, arising from the competition
between diffusive spreading, which increases entropy, and emergent
fine structure, which tends to decrease it.
This entropy maximum is a general property of systems driven by non-Gaussian discrete noise,
whose dynamics near stable fixed points can be viewed as an autoregressive
process -- an approximation that is mathematically equivalent to
unbiased random walks with shrinking steps. We discuss potential implications of Bernoulli convolution dynamics for protocell
self-replication and vesicle proliferation, establishing a link between our information-theoretic approach and
biophysical models of cell division.
\end{abstract}

\maketitle

\section{Introduction}


Entropy maximization, as postulated in the second law of thermodynamics, determines equilibrium states across a broad range of physical systems.
The appearance or disappearance of entropy maxima marks profound changes in macroscopic behavior such as phase transitions.
The local structure of entropy near its maxima defines thermodynamic forces and governs
near-equilibrium dynamics, in which these forces compete with stochastic diffusion driven by thermal noise
\cite{LandauLifshitz1980StatPhys1,Chandler1987}.

Random walks provide a natural framework for modeling diffusive behavior in physical and chemical systems~\cite{Weiss1994,Metzler2000,Klafter2011}.
Microscopic dynamics govern the statistics of individual steps, which in turn determine probability distributions over random walk endpoint positions
and thus the entropy. This entropy characterizes stable states, transitions between states, and can be used to derive thermodynamic forces. For certain
distributions -- particularly fractal ones -- entropy calculations are technically challenging and, in general, depend on the fractal resolution scale~\cite{Metzler2000}.

Random walks with shrinking steps~\cite{Krapivsky2004} can be used to describe the dynamics of physical systems in the vicinity of their stable states~\cite{Yeomans1992}.
These dynamics are modeled by the autoregressive process $x_{n+1}=rx_n+\zeta_n$, where
$x_n$ is the system state, $r \in (0,1)$ is the per-step contraction ratio (decay factor), and $\zeta_n$ is random noise. Iterating the
recursion gives
$x_n=r^{n}x_0+\sum_{k=0}^{n-1}r^{k}\zeta_{n-1-k}$, where
the sum represents a random walk with the contraction factor $r$~\cite{Sornette1998}. Relevant physical examples include optical line
broadening~\cite{Barkai1999b}, diffusion in sheared flows~\cite{Ben-Naim1992}, damage accumulation in
cells~\cite{Pikovsky2023}, and cell size regulation in bacteria~\cite{Amir2014}.

In particular, the size distribution of self-replicating cells with a constant average
size can be modeled as an autoregressive process~\cite{Amir2014,Elgamel2024},
equivalent to a diffusive process with geometrically shrinking steps and stochastic
noise. In this framework, the contraction ratio $r$
encodes the cell-size control strategy; for instance, $r=1/2$
corresponds to the adder model, in which a cell adds an approximately
constant volume before division. The noise $\zeta_n$ is typically sampled from a
Gaussian distribution, but it may also be non-Gaussian depending on the underlying molecular mechanisms~\cite{Elgamel2024}.
For example, in protocell models~\cite{Schrum2010} and in systems exhibiting Ostwald
ripening~\cite{Zwicker2017} or membrane instabilities with a
characteristic scale~\cite{Shivers2025}, the noise may be sampled from a discrete distribution, since cell growth proceeds by
engulfing other vesicles or droplets of a specific limiting size~\cite{Huang2017}.

Here, we show that the Shannon information-theoretic entropy of the probability distributions generated by unbiased one-dimensional random walks with shrinking steps (known as Bernoulli convolutions~\cite{Varju2016})
exhibits a local maximum at $r=1/2$. This maximum emerges above a certain threshold resolution and its basin of attraction shrinks at finer scales.
We discuss the emergent local entropy maximum in the context of a toy model of protocell division. Specifically, we introduce an adder-like
self-replicating model in which a cell can grow by one or two fixed-volume increments prior to division. This model naturally maps to the
Bernoulli convolution process and therefore yields the maximum-entropy size distribution in the special case of $r = 1/2$.
For polydisperse systems, such as colloids or early-life protocell populations, the size distribution is determined by the underlying growth and duplication dynamics.
There is both experimental and theoretical evidence that the entropy of size distributions contributes
significantly to the total free energy of polydisperse systems~\cite{Sollich2002}, influences duplication processes~\cite{Corominas-Murtra2019},
and affects the quality of colloidal crystals~\cite{Pusey1987,Arif2025}.

\section{Preliminaries}

Bernoulli convolutions model diffusion-like processes with discrete steps characterized by geometrically decaying magnitudes~\cite{Varju2016}.
Throughout this paper, we parameterize the per-step contraction ratio by $\lambda\equiv r^{-1}>1$, so that $r=\lambda^{-1}$.
The dyadic point corresponds to $\lambda=2$ ($r=1/2$). The probability distribution $P(x; \lambda,l)$ is the distribution of the endpoints $x$ of a symmetric (unbiased) random walk
whose step $k$ has a magnitude $\lambda^{-k}$:
\begin{equation}
x = \sum_{k=1}^{l} c_k \lambda^{-k}, \quad c_k \in \{-1, +1\},
\label{eq:bc_def}
\end{equation}
where $\lambda>1$ is the inverse of the per-step contraction ratio, each coefficient $c_k$ equals $-1$ or $+1$ with
probability $1/2$, and $l$ is the number of steps. Note that the sum
is bounded for $\lambda>1$ in the $l \rightarrow \infty$ limit, with $x \in
[-(\lambda-1)^{-1},(\lambda-1)^{-1}]$. Thus, Bernoulli convolutions yield bounded distributions that typically exhibit
non-trivial fractal structure. The fractal structure is due to self-similarity: in the $l \rightarrow \infty$ limit, after $l'$ steps, the remaining sum in Eq.~\eqref{eq:bc_def}
is a copy of the full distribution rescaled by $\lambda^{-l'}$. 

The entropy of Bernoulli convolutions has remained a largely open problem ever since Erd\H{o}s raised questions about the statistical properties of random series with contracting coefficients in the 1930s~\cite{Erdos1939}. Closed-form expressions for Bernoulli-convolution distributions are known only for a few special values of $\lambda$.
For most values of $\lambda$, the distributions are complex and counterintuitive, complicating both numerical computation and analytical derivations.
The entropy of Bernoulli convolutions is known in a few special cases admitting closed-form expressions,
including $\lambda=2^{k}$~\cite{Krapivsky2004} (smooth distribution) and the golden ratio $\lambda={(1+\sqrt{5})}/{2} \approx 1.618$ (fractal
distribution)~\cite{Alexander1991}. Entropy limits have also been investigated in connection with the number
theoretic properties of $\lambda$ (Pisot or Salem numbers)~\cite{Peres2000}.

Here, we compute the derivatives of the Bernoulli convolution entropy
in the neighborhood of the dyadic point $\lambda=2$.
Because the density distributions become fractal and discontinuous near this point, we first define the derivatives at a finite
coarse-graining scale and then take the vanishing-scale limits. We develop an explicit overlap-counting analytical approach inspired by classic estimators of entropy~\cite{KozachenkoLeonenko1987}
and mutual information~\cite{Grassberger1983} (an alternative formulation based on the continuity equation is presented in Appendix~\ref{app:continuity}).
Specifically, we equip a discrete set of endpoints of $l$-step Bernoulli random walks with tail kernel distributions.
We employ three independent numerical schemes to validate our analytical predictions near $\lambda=2$ and to study the behavior of the Bernoulli convolution entropy in a wider range of $\lambda$ values.

\section{Methods}

\subsection{Coarse-grained probability distribution}
\label{sec:prob}

\begin{figure}[htbp]
\centering
\includegraphics[width=.9\linewidth]{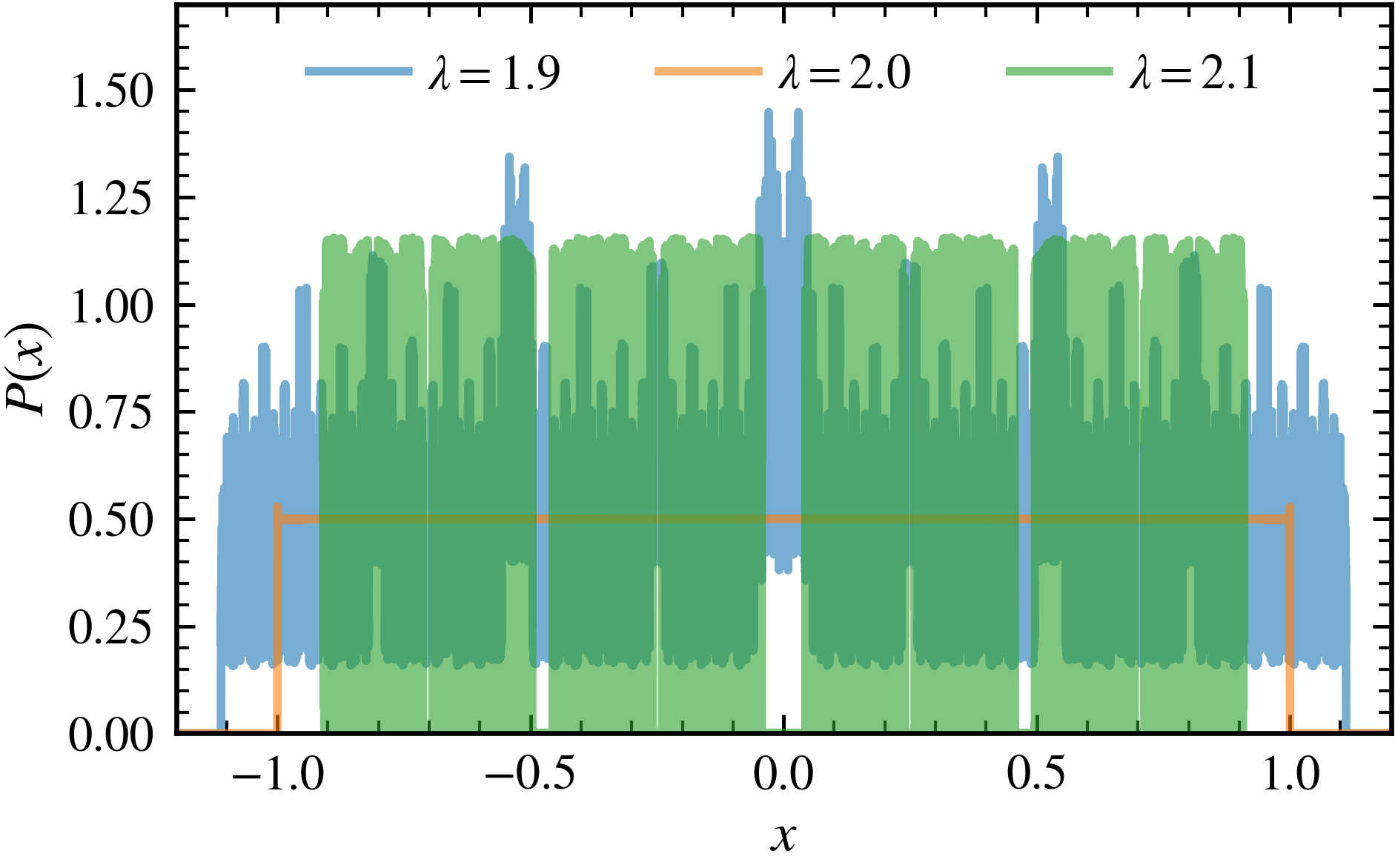}
\caption{\textbf{Probability distributions for random walks with
    shrinking steps (Bernoulli convolutions) near $\boldsymbol{\lambda = 2}$.} Coarse-grained densities $P(x;\lambda,l)$ at
  $\lambda=1.9$ (blue), $\lambda=2.0$ (red), and $\lambda=2.1$ (green)
  as functions of the coordinate $x$, computed using the boundary-tracking method with $l = 20$ (Methods; Appendix~\ref{app:numerics}).
  At $\lambda=2$, the distribution becomes uniform in the $l \rightarrow \infty$ limit, maximizing the entropy on its support. 
  For $\lambda=1.9$ (i.e., $\lambda<2$), the distribution broadens and develops complicated internal structure. 
  For $\lambda=2.1$ (i.e., $\lambda>2$), the distribution contracts and becomes increasingly disconnected due to the gaps between adjacent regions with non-zero probability density.
} 
\label{fig:fig-dist}
\end{figure}

Figure~\ref{fig:fig-dist} shows the behavior of $P(x; \lambda,l)$,
the probability distribution generated by Bernoulli convolutions
(Eq.~\eqref{eq:bc_def}) near $\lambda = 2$. At $\lambda=2$, $P(x; \lambda,l)$ is rectangular in the $l \rightarrow \infty$ limit:
$\lim_{l \rightarrow \infty} P(x; 2, l)=1/2$ for $|x|<1$ and zero otherwise. This follows directly from Eq.~\eqref{eq:bc_def}, which
yields $2^{l}$ uniformly distributed endpoints after $l$ steps. The set of endpoints tiles the $[-1,1]$ interval in the $l \rightarrow \infty$ limit. As $\lambda$ deviates from $2$, two trends emerge: (i) the distribution broadens for $\lambda<2$ and contracts for $\lambda>2$; and (ii) a pronounced internal structure develops, possibly fractal. For $\lambda>2$, the contraction splits the distribution into isolated regions separated by gaps.

Figure~\ref{fig:fig-dist} illustrates the trade-off between
broadening of the distribution, which increases entropy, and structure formation, which
decreases it. At $\lambda = 2$, the distribution converges to uniform on the $[-1,1]$ interval as the number of steps increases, 
maximizing the entropy.
When $\lambda<2$, the distribution
broadens, increasing the entropy, while any emergent structure
tends to decrease it; the net change in entropy depends on which effect
dominates. When $\lambda>2$, the distribution narrows and gaps appear;
both effects lower the entropy, leading to its overall
decrease.

For a generic $\lambda>1$, the limiting Bernoulli–convolution measure is often singular or fractal, so that the continuous entropy is undefined.
We therefore compute a coarse-grained Shannon entropy at a given spatial resolution $l$, in the spirit of neighborhood-based estimators of probability density~\cite{KozachenkoLeonenko1987,Grassberger1983}.

Specifically, we define resolution $l \geq 1$ as $l$ explicit Bernoulli convolution steps, with the rest
of the steps represented by a tail-kernel estimator. We keep track of
the $2^l$ terms corresponding to all distinct trajectories of length $l$ in Eq.~\eqref{eq:bc_def}, with
the endpoints $x_i^{(l)}~(i = 1 \dots 2^l)$. 
This procedure leaves unresolved the tail of the distribution (with $k > l$ steps), which has a half-width of
\begin{equation}
a_l(\lambda)=\sum_{k=l+1}^{\infty}\lambda^{-k}=\frac{1}{\lambda^{l}(\lambda-1)}. 
\label{eq:tail_halfwidth}
\end{equation}
We approximate the unresolved tail by a uniform (rectangular, or \texttt{rect}) kernel. This
approximation is justified in the vicinity of $\lambda=2$ because the Bernoulli convolution distribution
is a \texttt{rect} function at $\lambda=2$ (Fig.~\ref{fig:fig-dist}). Consequently, each
endpoint $x_i^{(l)}$ is replaced by a rectangular (uniform) tail kernel defined over the $[x_i^{(l)}-a_l(\lambda),\,x_i^{(l)}+a_l(\lambda)]$ interval with the density
\begin{equation}
h(\lambda)=\frac{2^{-l}}{2a_l(\lambda)}=\frac{\lambda^{l}(\lambda-1)}{2^{l+1}}.
\label{eq:tail_density}
\end{equation}
Thus, the coarse-grained density $P(x;\lambda,l)$ is the sum of $2^l$
\texttt{rect} functions that may overlap with one another.
In other words, each $l$-step trajectory $i$ yields a partial endpoint position $x_i^{(l)}$,
while the remaining steps contribute an additional displacement confined to $x_i^{(l)} \pm a_l(\lambda)$.
We model this tail displacement as a uniform distribution on $[x_i^{(l)}-a_l(\lambda),\,x_i^{(l)}+a_l(\lambda)]$, so that each trajectory contributes a rectangular ``patch'' of width $2 a_l (\lambda)$ and weight $2^{-l}$.
Note that at $\lambda=2$, the patch width is $2a_l(2)=2^{-l+1}$, yielding $h(2)=\frac{1}{2}$ in Eq.~\eqref{eq:tail_density}. Eq.~\eqref{eq:tail_halfwidth} shows that
larger $l$ yields finer patches. Crucially, exact patch widths are retained when forming $P(x;\lambda,l)$, such that the widths of the overlaps between neighboring patches also remain exact;
only the patch-scale probability density structure is smoothed by making it uniform.

\subsection{Overlap-counting entropy derivation}
\label{sec:overlap}

The Shannon entropy corresponding to $P(x;\lambda,l)$ is defined as
\begin{equation}
S(\lambda,l)=-\int_{-\infty}^{\infty} P(x;\lambda,l)\,\ln P(x;\lambda,l)\,dx.
\label{eq:entropy_def}
\end{equation}
We use it primarily as an information-theoretic measure which characterizes the structure of the coarse-grained probability distribution $P(x;\lambda,l)$ at a given resolution $l$.
Our goal is to evaluate the entropy and its first derivative with respect to $\lambda$ in the vicinity of $\lambda = 2$.

To evaluate $S(\lambda,l)$
at $\lambda=2$, we note that all contributions to $P(x;\lambda,l)$
from steps $k > l$ are rectangular and have the height of $1/2$, see
Eqs.~\eqref{eq:tail_halfwidth} and \eqref{eq:tail_density}. These rectangles pack tightly
and merge into a single large rectangle. Consequently, the behavior of the entropy
near $\lambda=2$ is asymmetric: for $\lambda<2$ the $P(x;\lambda,l)$ distribution is
formed by overlapping rectangles, whereas for $\lambda>2$ the
rectangles are separated by gaps (Fig.~\ref{fig:fig-dist}). As a result, the left and right derivatives of the entropy must be
computed separately.

\begin{figure}[t]
\centering
\includegraphics[width=0.9\linewidth]{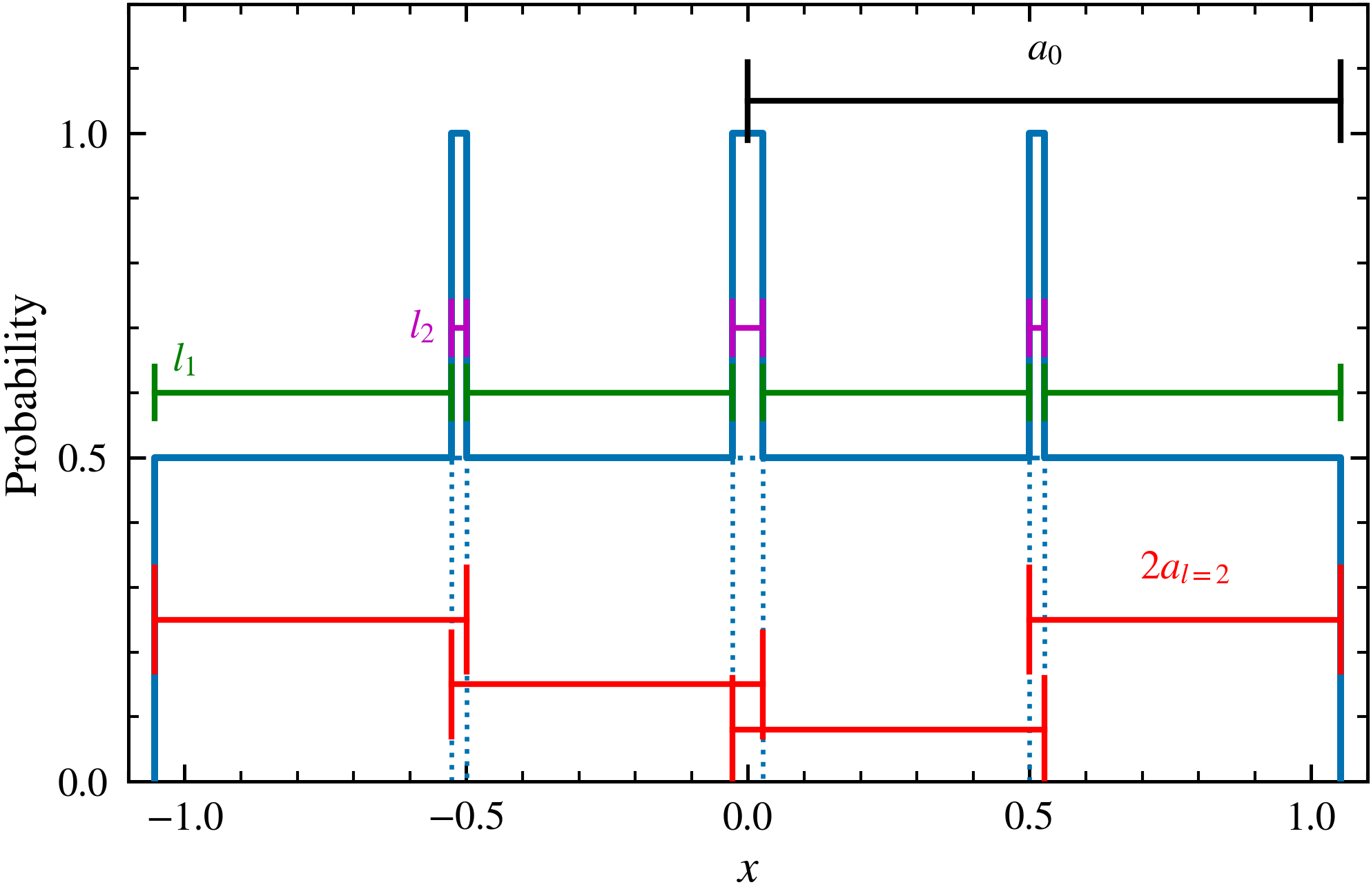}
\caption{\textbf{Coarse-grained structure of the probability density.}
  Shown is $P(x;\lambda,l)$ for $\lambda=1.95$ and $l=2$ (blue lines). Each of the
  $2^2=4$ paths contributes a rectangular kernel of width $2 a_l$ (Eq.~\eqref{eq:tail_halfwidth}; red
  lines with bars). For $\lambda<2$, adjacent kernels overlap,
  creating regions with density $\approx 1/2$ (single coverage, total
  length $\ell_1$, green lines with bars) and $\approx 1$ (double coverage,
  total length $\ell_2$, magenta lines with bars). The half-width of the full
  distribution is $a_0=(\lambda-1)^{-1}$ (black arrow with bars). The piecewise structure of $P(x;\lambda,l)$ enables analytical
  entropy calculation via Eqs.~\eqref{eq:length_constraints}--\eqref{eq:entropy_below} and is also exploited in the boundary-tracking numerical method
  (Methods; Appendix~\ref{app:numerics}).
  }
\label{fig:Px_schematic}
\end{figure}

\textbf{Case $\lambda < 2$ (overlap regime).} The weight of each path with $l$ steps is $2^{-l}$, and the corresponding tail density for each of the $2^{l}$ rectangular tail kernels is $h(\lambda)$ given by Eq.~\eqref{eq:tail_density}.
For $\lambda$ just below 2, adjacent rectangles overlap once, creating regions with density $h$ (single coverage) and $2h$ (double coverage).
We define $\ell_{1}(\lambda)$ as the total length of regions with density $h$ and $\ell_{2}(\lambda)$ as the total length of regions with density $2h$. These lengths satisfy
\begin{equation}
\begin{cases}
\ell_{1}(\lambda) + \ell_{2}(\lambda) = 2 a_{0}(\lambda),\\
\ell_{1}(\lambda) + 2\,\ell_{2}(\lambda) = 2^{\,l+1} a_{l}(\lambda),
\end{cases}
\label{eq:length_constraints}
\end{equation}
where $a_0(\lambda) = (\lambda-1)^{-1}$ is the half-width of the full
distribution; see Figure~\ref{fig:Px_schematic} for a representative example. 

Equation~\eqref{eq:length_constraints} yields:
\begin{equation}
\begin{aligned}
\ell_1(\lambda) &= a_l(\lambda) \bigl[ 4\lambda^l - 2^{l+1} \bigr],\\
\ell_2(\lambda) &= a_l(\lambda) \bigl[ 2^{l+1} - 2\lambda^l \bigr].
\end{aligned}
\end{equation}
The entropy is given by (Eq.~\eqref{eq:entropy_def}):
\begin{equation}
S(\lambda,l) = -\ell_1 h \ln h - \ell_2 (2h) \ln(2h),
\end{equation}
which simplifies to:
\begin{equation}
S(\lambda,l)
= \left[ \frac{\lambda^l}{2^{l-1}} + (l-1) \right] \ln 2 - l\ln\lambda - \ln(\lambda-1).
\label{eq:entropy_below}
\end{equation}
Consequently,
\begin{equation}
\left.\frac{\partial S}{\partial \lambda}\right|_{2^-}
= l\!\left(\ln 2 - \tfrac{1}{2}\right) - 1.
\label{eq:dS_left}
\end{equation}
The left derivative in Eq.~\eqref{eq:dS_left} changes sign at
\begin{equation}
l_{\mathrm{th}} = \frac{1}{\ln 2 - 1/2} \approx 5.18.
\label{eq:threshold}
\end{equation}

\textbf{Case $\lambda > 2$ (gap regime).} The rectangles no longer overlap in this case; instead, gaps of zero density appear. Hence $\ell_{2}=0$ and the entropy is determined solely by $\ell_{1} = 2^{l+1} a_l(\lambda)$:
\begin{equation}
S(\lambda,l) = \ln\frac{2}{\lambda-1} - l\,\ln\frac{\lambda}{2}.
\label{eq:entropy_above}
\end{equation}
Differentiation yields 
\begin{equation}
\left.\frac{\partial S}{\partial\lambda}\right|_{2^+} = -\frac{l}{2} - 1.
\label{eq:dS_right}
\end{equation}
The right derivative in Eq.~\eqref{eq:dS_right} is always negative.

For $l > l_{\mathrm{th}}$
(i.e., $l > 5$ in practice), the left derivative is positive and
the right derivative is negative, implying a cusp-like local maximum
at $\lambda=2$. For $l \leq l_{\mathrm{th}}$, there is no local maximum since both derivatives are negative.

\subsection{Numerical algorithms for computing fractal entropy}

We implement three complementary numerical methods to estimate
$S(\lambda,l)$ and validate our analytical results: 
boundary tracking (BT), Fast Fourier Transforms (FFT), and grid-based iteration (GBI).
Each method is briefly introduced below;
see Appendix~\ref{app:numerics} for detailed descriptions.

\begin{enumerate}
\item \textbf{BT:}
At resolution $l$, enumerate all $2^l$ paths, compute their endpoint positions, and replace each endpoint by a rectangular tail kernel of half-width $a_l(\lambda)$. Sort all interval boundaries and count overlaps to obtain the piecewise-constant probability density. Compute entropy by summation over all constant-probability segments. 

\item \textbf{FFT:} Evaluate the characteristic function of the truncated Bernoulli convolution on a uniform $k$-grid, apply inverse FFT and coarse-grain the probability distribution
to the target bin size $\Delta x = 2^{1-l_\text{FFT}}$. Compute entropy by summation over all $\Delta x$ bins.

\item \textbf{GBI:} Iterate the random walk (transfer) operator $p \mapsto \frac{1}{2}[T_{+} p + T_{-} p]$, where $T_\pm(x) = (x \pm 1)/\lambda$,
on a fixed grid with bin size $\Delta x = 2^{1-l_\text{GBI}}$, until the probability distribution converges. Compute entropy by summation over all $\Delta x$ bins.
\end{enumerate}

The BT method uses the same approximations as our analytical calculations and therefore becomes asymptotically exact in the vicinity of $\lambda=2$, where the tail kernels are closely approximated
by the \texttt{rect} function. As $\lambda$ deviates from $2$, both the BT numerical calculation and the theory become less accurate since tail kernels should adopt the self-similar form
of the general distribution in Eq.~\eqref{eq:bc_def}, which ceases to be a \texttt{rect} function.

The FFT and GBI methods do not rely on the assumption of uniform tail kernels and therefore
provide independent numerical verification of our findings. Since it is numerically challenging to compute the entropy of fractal probability distributions and
the numerical calculations are sensitive to the spatial resolution (which is implemented differently by each method, see Appendix~\ref{app:numerics}), our goal is not to match the numerical results exactly, but to show
that all three computational methods support the emergence of an entropy maximum at $\lambda=2$.

\section{Results}

\begin{figure}[htbp]
\centering
\includegraphics[width=.9\linewidth]{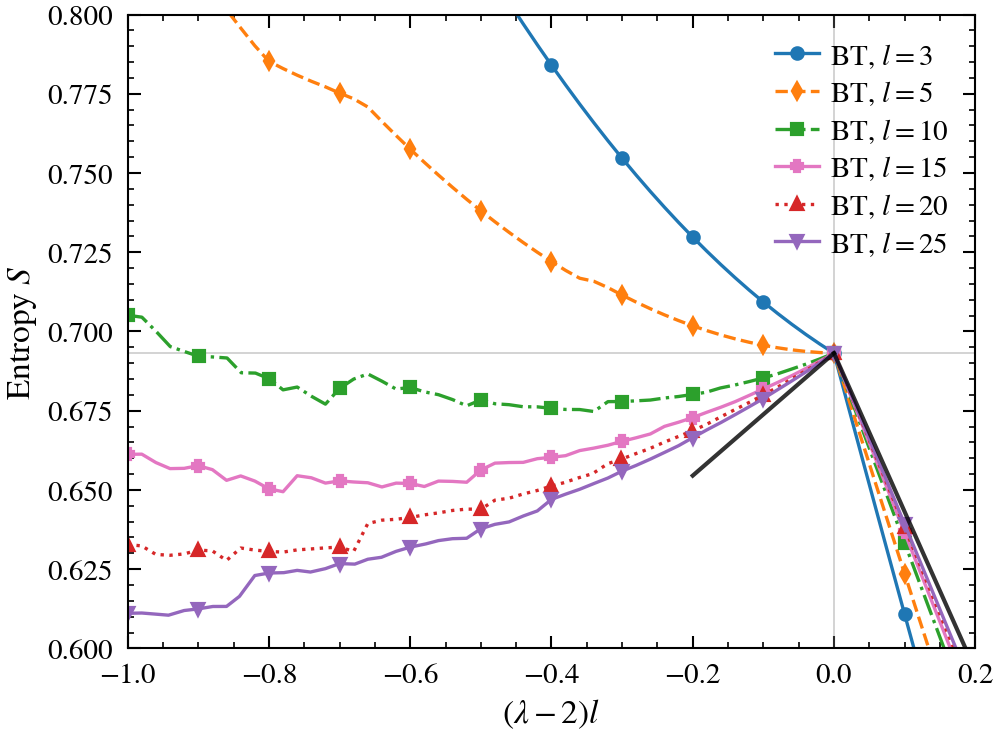}
\caption{
\textbf{Fractal entropy in the vicinity of $\lambda=2$.}
Coarse-grained entropy $S(\lambda,l)$ was computed using BT (Methods; Appendix~\ref{app:numerics}) and plotted versus $(\lambda-2)l$ for several resolution levels $l \in \{3,5,10,15,20,25\}$.
A local maximum emerges for $l \geq 6$. Black lines show the asymptotic slopes given by Eqs.~\eqref{eq:limit_left} and \eqref{eq:limit_right} -- see Eq.~\eqref{eq:taylor_expansion}.
Each numerical curve is based on $101$ independent entropy calculations: $(\lambda-2)l = (-1.00, -0.98, \dots, 0.98, 1.00)$.
}
\label{fig:fig-univer}
\end{figure}

Our central analytical result is the emergence of a local maximum of the fractal
entropy of Bernoulli convolutions at $\lambda=2$. The existence of this maximum follows from the
one-sided derivatives of the entropy at $\lambda=2$: $\left. {\partial S}/{\partial \lambda}\right|_{2^-} > 0$ for $l \ge 6$ (Eq.~\eqref{eq:dS_left}),
whereas $\left. {\partial S}/{\partial \lambda}\right|_{2^+} < 0$ (Eq.~\eqref{eq:dS_right}).
Figure~\ref{fig:fig-univer} shows the behavior of the fractal entropy $S(\lambda,l)$ in the vicinity of $\lambda = 2$, plotted as a function of $\Delta \lambda\,l$, where $\Delta \lambda \equiv \lambda - 2$.
We find empirically that this coordinate manifests the universal behavior of the entropy curves plotted at different $l$: $\Delta\lambda \sim 1/l$.
Overall, we observe an excellent agreement between the BT numerical method (Methods;
Appendix~\ref{app:numerics}) and the analytically computed slopes of the entropy function, Eqs.~\eqref{eq:dS_left} and \eqref{eq:dS_right}. As predicted analytically,
a prominent local maximum appears in the numerical curves for $l > 5$.

Note that as $l \rightarrow \infty$, the normalized derivatives converge to finite limits:
\begin{equation}
\frac{1}{l}\frac{\partial S}{\partial\lambda}\bigg|_{2^{-},\,l\rightarrow\infty}=\ln 2 - \frac{1}{2}
\approx 0.193,
\label{eq:limit_left}
\end{equation}
and
\begin{equation}
\frac{1}{l}\frac{\partial S}{\partial\lambda}\bigg|_{2^{+},\,l\rightarrow\infty}=-\frac{1}{2}.
\label{eq:limit_right}
\end{equation}
Eqs.~(\ref{eq:limit_left}) and (\ref{eq:limit_right}) follow from Eqs.~(\ref{eq:dS_left}) and (\ref{eq:dS_right}), respectively.

We can plot these limiting values of entropy derivatives using a first-order Taylor expansion near $\lambda = 2$ (cf. black lines in Fig.~\ref{fig:fig-univer}):
\begin{equation}
   S \simeq S(2) + \left[ \frac{1}{l}\frac{\partial S}{\partial\lambda}\bigg|_{2^{\pm}} \right] \, l (\lambda-2).
  \label{eq:taylor_expansion}
\end{equation}

As resolution $l$ increases, the cusp-like maximum at $\lambda=2$ becomes sharper and less wide (Fig.~\ref{fig:fig-univer}). Empirically, its basin of attraction shrinks as $l$ increases, since
$|\Delta \lambda| \sim 1/l$. The practical relevance of the maximum depends on whether deviations in $\lambda$ are small enough to remain within its basin of attraction at a given resolution $l$. 

\begin{figure}[htbp]
\centering
\includegraphics[width=.9\linewidth]{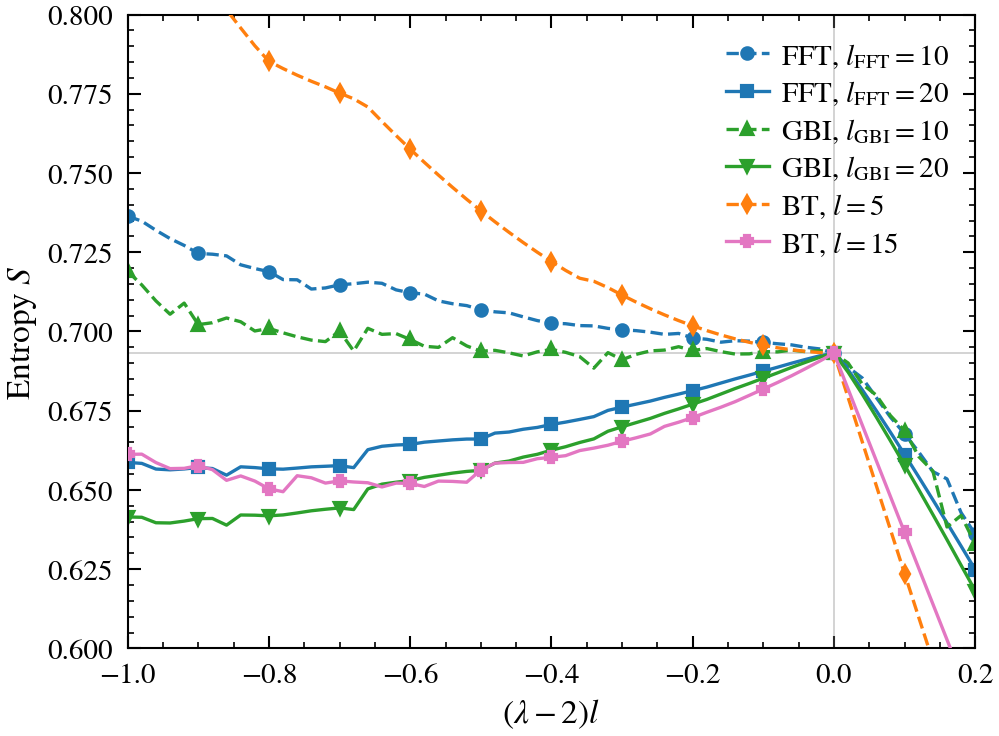}
\caption{ \textbf{Comparison of numerically computed fractal entropy curves.}
Shown are entropies of Bernoulli convolutions in the vicinity of $\lambda=2$,
computed using BT ($l=5,15$), FFT ($l_\text{FFT}=10,20$), and GBI ($l_\text{GBI}=10,20$).
The emergence of the local maximum at $\lambda=2$ is corroborated by all three numerical methods.
Each numerical curve is based on $101$ independent entropy calculations: $(\lambda-2)l = (-1.00, -0.98, \dots, 0.98, 1.00)$.
}
\label{fig:fig-SvsLam}
\end{figure}

To ensure that the observed behavior of the fractal entropy in the vicinity of $\lambda=2$ does not depend on the tail-kernel approximation used in the BT method and the analytical calculations,
we have computed the entropy using two other numerical methods, FFT and GBI (Methods; Appendix~\ref{app:numerics}) (Fig.~\ref{fig:fig-SvsLam}).
We find that all three numerical methods yield qualitatively similar entropy curves, which exhibit the emergence of the local entropy maximum at $\lambda=2$ above a critical resolution scale $l_\text{th}$.

The similarity becomes more pronounced at larger values of $l$ (cf. solid curves in Fig.~\ref{fig:fig-SvsLam}).
However, since resolution scales are implemented somewhat differently in the three methods (Appendix~\ref{app:numerics}), we find that $l_\text{th}$ computed by the BT method does not
correspond to $l_\text{th,FFT}$ and $l_\text{th,GBI}$. Empirically, we observe that setting $l_\text{th,FFT} \simeq l_\text{th,GBI} \simeq l_\text{th} + 5$ correctly matches the emergence of the local maximum at
$\lambda=2$, even though the shape of the curves at $\lambda < 2$ is better reproduced at $l \simeq l_\text{th,FFT} \simeq l_\text{th,GBI}$ (cf. $l=10$ and $l=20$ BT curves in Fig.~\ref{fig:fig-univer}).


Numerical calculations shown in Figs.~\ref{fig:fig-univer} and \ref{fig:fig-SvsLam} indicate that at any scale $l$, the entropy eventually starts increasing as $\lambda$ decreases.
This is a consequence of the distribution support growing unboundedly as $\lambda \to 1^+$. Thus, the maximum at $\lambda=2$ is local, with a finite basin of attraction.

To connect our results to the dynamics of cell growth and division, we introduce a
minimal ``binary-adder'' modification of the classic adder model of cell division~\cite{Amir2014}: the added
volume per cycle is drawn equiprobably from $\{\Delta v,2\Delta v \}$. Let $V_{n}$ denote the post-division cell volume at division $n$. The cell volume update rule is
\begin{equation}
V_{n+1}=\frac{V_n+\Delta_{n+1}}{2},\qquad \Delta_n\in\{\Delta v,2\Delta v\}.
\label{eq:recursion}
\end{equation}

The explicit finite-$n$ solution of Eq.~\eqref{eq:recursion} is given by
\begin{equation}
V_n=\frac{V_0}{2^n}+\frac{3}{2}\Delta v\,(1-2^{-n})+\frac{\Delta v}{4} \sum_{k=0}^{n-1} 2^{-k} \, c_{n-k},
\label{eq:recursion2}
\end{equation}
where $V_0$ is the initial volume of the cell and we used $\Delta_m = \tfrac{3}{2} \Delta v + \tfrac{1}{2} \Delta v \, c_m$, with $c_m \in \{-1,+1\}$.
Note that each set of independent random variables $\{ c_n, \dots, c_1 \}$ in the last term on the right-hand side of Eq.~\eqref{eq:recursion2} represents a specific random walk.
Since we are interested in the ensemble of all random walks, we can replace $\{ c_n, \dots, c_1 \}$ by $\{ c_1, \dots, c_n \}$ in the sum over $k$ -- this will only permute the paths in the ensemble,
leaving the ensemble itself invariant. Thus, $\sum_{k=0}^{n-1} 2^{-k} \, c_{n-k} \to \sum_{k=0}^{n-1} 2^{-k} \, c_{k+1} = 2 \sum_{k=1}^{n} 2^{-k} \, c_{k}$ in Eq.~\eqref{eq:recursion2}.

In the $n \to \infty$ limit, we obtain:
\begin{equation}
V_\infty = \frac{3}{2} \Delta v + \frac{\Delta v}{2} \sum_{k=1}^{\infty} c_k\,2^{-k},
\label{V:inf}
\end{equation}
which corresponds to a Bernoulli convolution (Eq.~\eqref{eq:bc_def}) with $\lambda=2$, centered at $3\Delta v/2$ and defined on $[\Delta v,2\Delta v]$.

More generally, if cell division yields a fraction $1/\lambda$ of the pre-division sum, $V_{n+1}=(V_n+\Delta_{n+1})/\lambda$ with $\lambda>1$, Eq.~\eqref{V:inf} generalizes to:
\begin{equation}
V_\infty(\lambda) = \frac{3\Delta v}{2(\lambda-1)} + \frac{\Delta
  v}{2} \sum_{k=1}^{\infty} c_k\,\lambda^{-k},
\label{V:inf:gen}
\end{equation}
which corresponds to the Bernoulli convolution with parameter $\lambda$ centered at $3\Delta v/[2(\lambda-1)]$.


\begin{figure}[htbp]
\centering
\includegraphics[width=.9\linewidth]{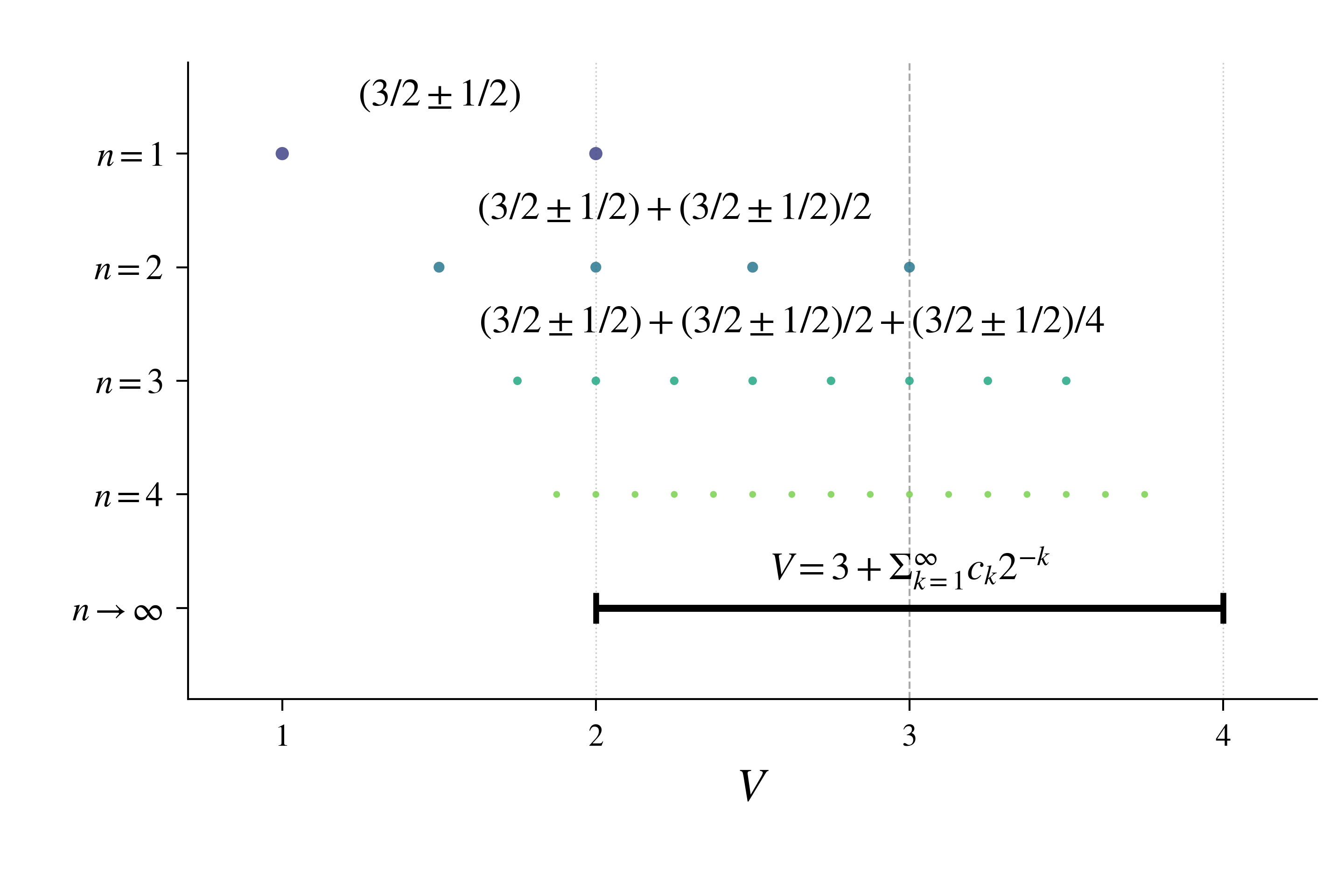}
  \caption{\textbf{Cell growth and division model with binary volume increments as a dyadic Bernoulli convolution.}
  Shown are the spectra of $2^n$ cell volumes after $n=\{ 1,2,3,4,\infty \}$ cell division events (Eq.~\eqref{eq:recursion}), with $\Delta v=2$ and $V_0=0$. At a given $n$, each point in the spectrum
  corresponds to an endpoint of a random walk with $n$ steps. In the $n \to \infty$ limit, the horizontal black bar shows the support of $V_\infty$ in Eq.~\eqref{V:inf}.
  }
\label{fig:cell-diagram}
\end{figure}

\section{Discussion and Conclusion}


In this work, we have studied unbiased random walks in which step sizes decay geometrically -- each subsequent step is smaller than the previous one by a fixed contraction ratio $r$.
Random walks with shrinking steps exhibit self-similarity and fractal behavior~\cite{Krapivsky2004}; consequently, 
their properties depend on the number of steps $l$, which sets the level of fractal resolution.
The endpoints of symmetric random walks with $l$ progressively smaller steps are given by the Bernoulli convolution, Eq.~\eqref{eq:bc_def}.
In this work, we focus in particular on computing the Shannon information-theoretic entropy of the ensemble of random walks, Eq.~\eqref{eq:entropy_def}.

In order to describe the limiting behavior of shrinking-step random walks analytically, we have developed the tail-kernel approximation, which replaces detailed random walk probability distributions after $l$ explicit steps
with rectangular kernels. This approximation, inspired by probability density estimation techniques~\cite{Bishop_2006}
and by classic estimators of entropy~\cite{KozachenkoLeonenko1987} and mutual information~\cite{Grassberger1983},
is exact at the dyadic point ($r=1/2$) and becomes progressively less valid away from it, since the probability density develops an internal
structure which is no longer correctly described by the uniform distribution.

We call the Shannon entropy of $l$-step random walks augmented with tail kernels the coarse-grained entropy of a Bernoulli convolution.
We find that the coarse-grained entropy exhibits a cusp-like local maximum at the dyadic point $r=1/2$ when the fractal resolution
exceeds $l = 5$ steps. The entropy maximum arises due to the competition between diffusive broadening, which increases entropy, and self-similar structure formation.
We observe that the width of the entropy maximum shrinks as $l$ increases, so that the maximum becomes sharper but narrower.
This behavior is confirmed qualitatively by two independent numerical techniques which do not rely on the tail kernel approximation: Fast Fourier Transforms and Grid-based Iterations (Methods; Appendix~\ref{app:numerics}).

Our findings are relevant to physical systems where the dynamics exhibit shrinking-step structure with discrete growth increments, such as the toy protocell division model in Eq.~\eqref{eq:recursion}, which maps directly
onto a Bernoulli convolution. Examples of discrete growth in biophysical systems include fusion events before instability-driven fission in vesicles or liquid droplets, mode-switching between low and high metabolic uptake states, and discretized cellular material uptake in early-life or minimal protocell settings.
Our results can be used to understand thermodynamics of such biophysical systems -- for example, the entropy of polydisperse size distributions is known to contribute to the total free energy in colloidal systems~\cite{Sollich2002,Ozawa2017}.
We note however that we do not develop detailed statistical mechanics models of colloidal or self-replicating systems in this work.

The autoregressive process described by Eq.~\eqref{eq:recursion} connects naturally to the problem of noise inheritance across generations of dividing cells.
The contraction ratio $r$ determines how ancestral fluctuations propagate through the successive rounds of cell division.
The $r=1/2$ value corresponds to the division of the mother cell into two equal parts and corresponds to geometric dilution across generations, consistent with the adder model of cell-size homeostasis~\cite{Amir2014,Elgamel2024} in which cells add an approximately constant volume before dividing, independent of their birth size. Thus, for $r=1/2$ the noise contribution from the cell's ancestors is simply divided by $2$ at each division cycle -- it is neither amplified nor actively suppressed. For $r<1/2$, inherited deviations are suppressed beyond simple dilution and the system forgets its history faster than geometric decay. For $r>1/2$, inherited deviations are reinforced and ancestral noise accumulates across generations. Although maintaining $r=1/2$ is likely to require active metabolic regulation, our entropy calculations suggest an intrinsic preference for symmetric division, at least  within the Bernoulli convolution framework.

Our analytical results rely on several specific features of the Bernoulli convolution. First, we consider unbiased random walks: each subsequent step goes left or right with equal probability;
asymmetric probabilities or multi-valued step distributions would modify the overlap and gap structure of the probability density $P(x;\lambda,l)$ and hence affect the entropy.
Second, step sizes decay geometrically: fluctuating or time-dependent contraction ratios may break the self-similarity exploited in our analysis.
Third, random walks exhibit self-similar branching: the resulting recursive structure allows us to introduce the tail-kernel approximation. 

In summary, our results illustrate how coarse-grained Shannon entropy of the Bernoulli convolution can exhibit counterintuitive behavior: increasing observation precision (quantified using $\Delta x \sim 2^{-l}$)
reveals probability density structures that reduce apparent randomness
away from the dyadic point $r=1/2$, leading to the local entropy maximum. In future work, we will extend our approach to account for asymmetric random walks with shrinking steps of
multiple sizes. We will also connect our treatment of the information-theoretic entropy with thermodynamic potentials and forces in the context of biophysical models of protocell division and
vesicle self-replication and proliferation.


\section*{Code and Data Availability}
The code and data required to reproduce all Figures in this paper are available at \url{https://github.com/afeigel/bernoulli-convolution-entropy}.

\section*{Acknowledgments}
A.V.M. acknowledges financial and logistical support from the Center for Quantitative Biology, Rutgers University.
A.F. is grateful to the Interdisciplinary Computational Physics Laboratory at the Racah Institute of Physics for the allocation of computational resources and technical assistance.


\appendix

\section{Continuity-equation calculation of the fractal entropy derivatives}
\label{app:continuity}

Here we present an alternative derivation of
Eqs.~\eqref{eq:dS_left} and~\eqref{eq:dS_right} by treating
$\lambda$ as an evolution parameter analogous to time and the
endpoint $x$ of each path as a particle. With this analogy, the
probability distribution $P(x; \lambda, l)$ satisfies a continuity equation:
\begin{equation}
\partial_{\lambda} P(x; \lambda, l)
  + \partial_{x} \bigl[ P(x; \lambda, l)\,v(x; \lambda, l) \bigr] = 0,
\label{eq:continuity}
\end{equation}
where $v(x;\lambda,l)$ is the velocity field, computed as the $\lambda$-derivative of the random walk endpoint
position $x_i^{(l)}(\lambda)$, with $i = 1 \dots 2^l$ labeling $l$-step random walk realizations. Each endpoint is augmented with a rectangular tail kernel,
as described in Section~\ref{sec:overlap}. In cases where two or more kernels overlap at a given position~$x$, the velocity is averaged over all overlapping kernels.

\begin{figure}[t]
  \centering
  \includegraphics[width=0.95\linewidth]{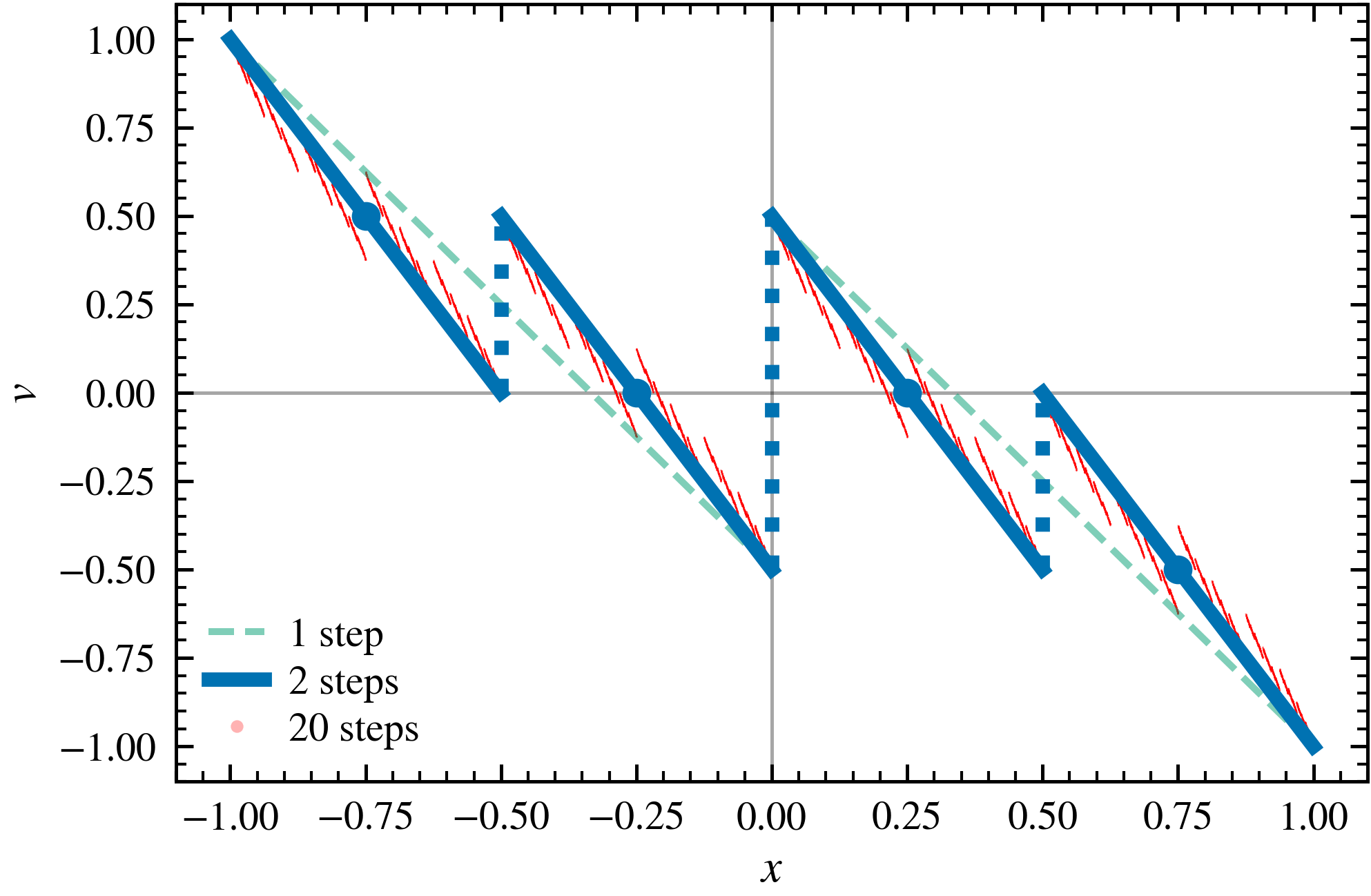}
  \caption{
\textbf{Velocity field in the continuity-equation framework.}
Velocity field $v(x;\lambda,l)$ for $\lambda\approx 2$ at several
resolutions: $l=1$ (teal), $l=2$ (blue), and $l=20$ (red). The velocity is
piecewise linear within each tail-kernel interval, with
discontinuities at the overlap boundaries where the tail kernels of the neighboring random walks meet
(compare with the structure of the probability density in Fig.~\ref{fig:Px_schematic}).
Blue circles on the $l=2$ curve mark the endpoints of four two-step random walks.
As $l$ increases, the velocity field becomes increasingly discontinuous and fine-grained.
}
  \label{fig:fig-SvsLam1}
\end{figure}

To compute $v(x;\lambda,l)$, we apply the same coarse-graining procedure to velocities as we did to
probability distributions (cf. Fig.~\ref{fig:Px_schematic}).  For each random-walk realization
$(c_1^{(i)},\ldots,c_l^{(i)})\in\{-1,+1\}^l$, the endpoint $x_i^{(l)}(\lambda) = \sum_{k=1}^l c_k^{(i)}\lambda^{-k}$
is extended beyond step~$l$ using a rectangular tail kernel of half-width~$a_l$ (Eq.~\eqref{eq:tail_halfwidth}).
The velocities at the left and right edges of each kernel are obtained by differentiating the corresponding
endpoint positions with respect to~$\lambda$. These velocities are interpolated
linearly across the extent of the kernel, yielding a piecewise-linear velocity field (Fig.~\ref{fig:fig-SvsLam1}).

Note that at $\lambda=2$, the velocities at the edges of the probability density support
$[-a_0,a_0]=[-1,1]$ are $v_{\mp 1} \equiv \mp\, da_0/d\lambda = \pm 1$
for all~$l$ (Fig.~\ref{fig:fig-SvsLam1}).  The slope of the velocity field within the $[-1,1]$ interval
depends on~$l$ and is determined by the widths of the tail-kernels.
Specifically, within each tail-kernel interval of width~$2a_l$ the slope is given by
\begin{equation}
  \label{eq:dvdx}
  \frac{\partial v}{\partial x}
    \simeq \frac{\Delta v(\lambda)}{a_{l}},
\end{equation}
where
\begin{equation}
\Delta v(\lambda) \equiv \frac{d a_l}{d\lambda}
  = -\sum_{k=l+1}^\infty k\,\lambda^{-k-1}
\label{eq:dv_tail_sum}
\end{equation}
is the change in boundary velocity across the half-width~$a_l$; this
corresponds to the Hausdorff derivative at resolution
$\Delta x = 2^{-l+1}$~\cite{Chen2006TimeSpaceFabric}.  The velocity
field is discontinuous at the boundaries between adjacent tail-kernel
intervals, with jumps of the magnitude~$\Delta \varv = v_R - v_L$,
where $v_R$ and $v_L$ refer to the velocity values at the right and left boundaries of the jump, respectively (Fig.~\ref{fig:fig-SvsLam1}).  At $\lambda=2$,
Eq.~\eqref{eq:dvdx} gives
\begin{equation}
\frac{\partial v}{\partial x} =
-\frac{l}{2} - 1,
\label{eq:dvdx_at_2}
\end{equation}
so that the slope is uniform across all tail-kernel intervals and depends
only on~$l$.


Next, we differentiate the endpoint position $x_i^{(l)}(\lambda)$ to obtain the velocity of path $i$:
\begin{equation}
v_i^{(l)} (\lambda)
  = -\sum_{k=1}^{l} k\,c_k^{(i)}\,\lambda^{-k-1}.
\label{eq:v_def}
\end{equation}
From Eq.~\eqref{eq:v_def}, the velocity difference between two
adjacent endpoints at $\lambda = 2$ is
\begin{equation}
\Delta v\big|_{\text{adjacent}} = 2^{-l}\,l.
\label{eq:delta_v_adj}
\end{equation}
At $l=2$, $\Delta v\big|_{\text{adjacent}} = 0.5$ is the difference
between the left and right blue circles that mark pairs of adjacent
endpoints in Fig.~\ref{fig:fig-SvsLam1}.

Each pair of adjacent endpoints, completed with tail kernels,
straddles a velocity discontinuity
(Fig.~\ref{fig:fig-SvsLam1}).  Subtracting the linear velocity
segments of width~$a_l$ on either side (with the slope given by
Eq.~\eqref{eq:dvdx_at_2}) yields the jump at a single
discontinuity:
\begin{equation}
\Delta \varv = -\!\left(\Delta v\big|_{\text{adjacent}}
  + 2\,\frac{\partial v}{\partial x}\,a_l\right) = 2^{1-l}.
\label{eq:v_single}
\end{equation}
Since there are $2^{l-1}$ discontinuities at the resolution level~$l$,
each level contributes $2^{l-1}\cdot 2^{1-l}=1$; summing over all
levels $l'=1,\dots,l$ gives
\begin{equation}
  \sum_{i} \Delta \varv_i = l,
  \label{eq:sum_dv}
\end{equation}
where the sum runs over all discontinuities. This can be verified graphically: for example, the heights of all discontinuities
(vertical dotted lines in Fig.~\ref{fig:fig-SvsLam1}) sum to~$l$ for
$l=1$ and $l=2$.

In the vicinity of $\lambda=2$, the probability distribution
$P(x;\lambda,l)$ is piecewise constant with overlaps
(Fig.~\ref{fig:Px_schematic}); the extent of the overlaps and the positions of the overlap boundaries depend on $\lambda$.
Thus, the total derivative of the entropy with respect to $\lambda$
must account for both the change in~$P(x;\lambda,l)$ and the motion of the overlap interfaces,
yielding, according to the Leibniz rule:
\begin{equation}
\label{eq:dS_general}
\begin{split}
\frac{d S}{d\lambda}
= -\int_{x \setminus \partial x}
    \partial_{\lambda} P(x; \lambda, l)\,
    \bigl[\ln P(x;\lambda,l)+1\bigr]\,dx \\
\quad + \sum_{\partial x_{i}} [P\ln P]_i\, v_{i},
\end{split}
\end{equation}
where the integral on the right-hand side is over the continuous regions (i.e., excluding
the discontinuities) and $v_i$ is the velocity at probability density discontinuity~$i$ (cf. Fig.~\ref{fig:Px_schematic} for the discontinuity positions and Fig.~\ref{fig:fig-SvsLam1} for the corresponding velocity values).
Finally, $[P\ln P]_i = (P\ln P)\big|_{i,\text{right}} - (P\ln P)\big|_{i,\text{left}}$ denotes
the jump of $P\ln P$ across interface (discontinuity)~$i$.
Near $\lambda=2$, the density takes the value $P=1/2$ over most of
the support, with narrow gaps ($P=0$) for $\lambda>2$ and narrow
overlaps ($P=1$) for $\lambda<2$; both vanish as $\lambda \to 2$.

Substituting Eq.~\eqref{eq:continuity} into
Eq.~\eqref{eq:dS_general} and integrating by parts decomposes the
entropy derivative into three contributions:
\begin{eqnarray} \label{eq:dS_decomposed}
\frac{d S}{d\lambda}
&=&  \int_{-\infty}^{\infty} \partial_{x} \left( v P \ln P \right) dx \\
&+& \int_{-\infty}^{\infty} P\,\partial_{x} v\,dx
+ \sum_{\partial x_{i}} [P\ln P]_i\, v_{i}. \nonumber
\end{eqnarray}
We label the three terms on the right-hand side~(A), (B), and~(C), respectively.

\textit{Term~(A)} reduces to boundary evaluations at the edges of each
tail kernel, since $\partial_x(vP\ln P)$ is a total derivative.
Within each $P=1/2$ region, $P \ln P = -\tfrac{1}{2} \ln 2$ is
constant, so that each kernel contributes
$-\tfrac{1}{2}\ln 2\cdot(v_R - v_L)$. Gaps and overlaps do not
contribute, since $P\ln P = 0$ at both $P = 0$ and $P = 1$; moreover,
these regions vanish as $\lambda\to 2$.  Thus, term (A) is given by the sum over the $P=1/2$ regions:
\begin{equation}
-\tfrac{1}{2}\ln 2\\
    \bigl[(v_{1}-v_{-1})
      - \textstyle\sum_i\Delta \varv_i\bigr]
  = (\tfrac{l}{2}+1) \ln 2,
\label{eq:A}
\end{equation}
where $v_{\pm 1} = \mp 1$ are the velocities at the support edges $x=\pm 1$
and $\Delta \varv_i$ is the velocity jump across gap~$i$; we used Eq.~\eqref{eq:sum_dv} in the final step.
Note that term~(A) is the same for $\lambda = 2^+$ and $\lambda = 2^-$.

\textit{Term~(B)} follows from Eq.~\eqref{eq:dvdx_at_2} and
$\int P\,dx = 1$:
\begin{equation}
  \int P\,\partial_{x}v\,dx = -\frac{l}{2} - 1.
  \label{eq:bulk_term}
\end{equation}
This term is also the same for $\lambda = 2^+$ and $\lambda = 2^-$.

\textit{Term~(C)} has to be calculated separately for outer (support) boundaries and
inner gaps or overlaps.  At the outer boundaries $x = \pm 1$, $P$
jumps between $0$ and $1/2$, resulting in
\begin{equation}
  v_{-1}\,P\ln P\big|_0^{1/2}
    + v_1\,P\ln P\big|_{1/2}^{0} = -\ln 2.
  \label{eq:C_outer}
\end{equation}
This value remains the same for $\lambda = 2^+$ and $\lambda = 2^-$.

At the inner boundaries, the $\lambda=2^-$ case produces overlaps (Fig.~\ref{fig:Px_schematic}), such that $P\!:\,1/2\!\to\!1\!\to\!1/2$.
Consequently, $v_R\,P\ln P\big|_{1/2}^{1} + v_L\,P\ln P\big|_{1}^{1/2} = +\tfrac{1}{2}\ln 2\,\Delta \varv$.
In the $\lambda=2^+$ case with gaps, $P\!:\,1/2\!\to\!0\!\to\!1/2$, resulting in
$v_L\,P\ln P\big|_{1/2}^0 + v_R\,P\ln P\big|_0^{1/2} = -\tfrac{1}{2}\ln 2\;\Delta \varv$.
Summing these contributions over all the inner interfaces, using Eq.~\eqref{eq:sum_dv},
and adding the outer-boundary contribution from Eq.~\eqref{eq:C_outer} yields
\begin{equation}
  - \ln 2 \pm \tfrac{l}{2} \ln 2
  \label{eq:C_total}
\end{equation}
for term (C), where the upper sign refers to $\lambda=2^-$ (overlaps) and the lower
to $\lambda=2^+$ (gaps).

Combining all three terms (Eqs.~\eqref{eq:A},~\eqref{eq:bulk_term}, and~\eqref{eq:C_total}), we obtain:
\begin{equation}
  \label{eq:total}
  \frac{dS}{d\lambda}\bigg|_{2^\pm}
  = -\frac{l}{2}-1
    +(\frac{l}{2} + 1) \ln 2
    -\ln 2
    \pm \frac{l}{2} \ln 2.
\end{equation}
The upper sign ($\lambda=2^-$) yields Eq.~\eqref{eq:dS_left}; for the lower sign ($\lambda=2^+$),
the logarithmic terms cancel, reproducing Eq.~\eqref{eq:dS_right}.

The continuity equation formulation clarifies the structure of 
Eq.~\eqref{eq:dS_decomposed}: the bulk term~(B),
$\int P\,\partial_x v\,dx$, captures the spreading effect, whereas the boundary term~(A)
captures the structure effect arising from density discontinuities.
The asymmetry between $\lambda=2^\pm$ limits resides entirely in the Leibniz
correction, term (C) (Eq.~\eqref{eq:C_total}).

\section{Numerical methods used in fractal entropy computations}
\label{app:numerics}

\subsection{Boundary Tracking}


\paragraph{Spatial resolution.}
The resolution parameter $l$ specifies the number of random walk steps, yielding $2^l$ distinct paths. Each path corresponds to a sequence of random variables $(c_1, c_2, \ldots, c_l) \in \{-1, +1\}^l$ and terminates at a position $x_i^{(l)}$ given by Eq.~\eqref{eq:bc_def}.

\paragraph{Algorithm.}
For each of the $2^l$ paths with endpoints $x_i^{(l)}$, we create a uniform tail kernel on an interval $[x_i^{(l)} - a_l (\lambda), x_i^{(l)} + a_l (\lambda)]$ (Eq.~\eqref{eq:tail_halfwidth})
with a height $h(\lambda)$ (Eq.~\eqref{eq:tail_density}). The probability density is piecewise constant, with a value proportional to the number of overlapping intervals at each point. We use a sweep-line algorithm to:
\begin{enumerate}
  \item Generate $2^{l+1}$ boundary events (starts and ends of $2^l$ tail kernel intervals).
  \item Sort all events by position.
  \item Sweep through the events, tracking the overlap count to construct the probability density.
\end{enumerate}
The entropy is computed by the exact summation over all piecewise constant segments:
\begin{equation}
  S = -\sum_{\text{segments } j} \Delta x_j \, P_j \ln P_j,
\end{equation}
where $\Delta x_j$ and $P_j$ are the width and the height of the segment $j$, respectively. Segments with $P_j < 10^{-14}$ were discarded from the sum.

The BT method is exact for the $l$-step truncated distribution. The terms with $k > l$ are approximated with rectangular kernels.



\paragraph{Computational complexity.}
The algorithm requires $\mathcal{O}(2^l)$ path enumerations and $\mathcal{O}(2^l l)$ operations for sorting the boundary events, making it feasible up to $l = 25$--$30$ on modern hardware.

\subsection{FFT Method}
\label{sec:fft_method}

The FFT method exploits the product structure of the characteristic function to compute the probability density efficiently. For the truncated Bernoulli convolution with $l$ terms, the characteristic function is
given by
\begin{equation}
  \hat{P}(k;\lambda) = \prod_{i=1}^{N} \cos(k\,\lambda^{-i}),
  \label{eq:char_function}
\end{equation}
where $N= 80$, ensuring convergence well beyond the resolution limit of our calculations.

\paragraph{Spatial resolution.}
The FFT calculation involves two distinct spatial resolutions:

\begin{enumerate}
\item \textbf{FFT resolution} $\Delta x_{\mathrm{fft}}$ sets the fine grid spacing at which the inverse FFT transform computes the probability density $P(x; \lambda,l_\text{FFT})$. This resolution is determined by the FFT array size in $k$-space, $N_{\mathrm{fft}}$, and the width of the computational window, $2 x_{\mathrm{s}}$:
\begin{equation}
  \Delta x_{\mathrm{fft}} = \frac{2 x_{\mathrm{s}}}{N_{\mathrm{fft}} - 1},
\end{equation}
where $x_{\mathrm{s}} = C_{\mathrm{w}} a_0$. Here, $a_0 = (\lambda-1)^{-1}$ is the half-width of the probability distribution support and
$C_{\mathrm{w}} = 50$ provides sufficient padding to minimize aliasing.

\item \textbf{Coarse-grained resolution} $\Delta x$ sets the bin width at which the entropy is computed. To match the BT method with $2^{l_\text{FFT}}$ bins over the support of width 2 (at $\lambda = 2$), we use:
\begin{equation}
  \Delta x = \frac{2}{2^{l_\text{FFT}}} = 2^{1-l_\text{FFT}}.
  \label{eq:delta:x}
\end{equation}
\end{enumerate}
We have ensured that $\Delta x_{\mathrm{fft}} \le \Delta x$ by setting $N_{\mathrm{fft}} = 2^{24}$ for $l_\text{FFT}=10$ ($\Delta x = 327.68 \, \Delta x_{\mathrm{fft}}$) and
$N_{\mathrm{fft}} = 2^{26}$ for $l_\text{FFT}=20$ ($\Delta x = 1.28 \, \Delta x_{\mathrm{fft}}$) (Fig.~\ref{fig:fig-SvsLam}).

\paragraph{Probability and entropy estimation.}
The fine-resolution density $P(x; \lambda, l_\text{FFT})$ computed at $\Delta x_{\mathrm{fft}}$ intervals is coarse-grained to the target resolution $\Delta x$ via cumulative distribution function (CDF) interpolation.
We first compute the CDF by summation of $P(x; \lambda, l_\text{FFT})$ at $\Delta x_{\mathrm{fft}}$ resolution, then take the difference between two CDF values at the edges of the coarse-grained bin $j$ of width $\Delta x$
to obtain the probability density in the bin, $P_j$. Finally, we compute the entropy as the sum over all bins:
\begin{equation}
  S = -\Delta x \sum_{\text{bins } j=1}^{2^{l_\text{FFT}}} P_j  \ln P_j.
  \label{eq:entropy:delta:x}
\end{equation}
As in the BT method, bins with $P_j < 10^{-14}$ were discarded from the sum.



\subsection{Grid-based Iteration}
\label{sec:grid_method}

The GBI method constructs the probability density by iterating the random walk (transfer) operator on a discrete grid with the fixed bin width $\Delta x$.

\paragraph{Spatial resolution.}
The probability distribution is discretized on a uniform grid spanning the $[-a_0, a_0]$ support interval, where $a_0 = (\lambda-1)^{-1}$ (Eq.~\eqref{eq:tail_halfwidth}).
The transfer operator is given by $T_\pm(x) = (x \pm 1)/\lambda$; it determines which source bins contribute to each target bin. The number of bins is
given by
\begin{equation}
  n_{\mathrm{bins}} = \left\lceil \frac{2 a_0}{\Delta x} \right\rceil,
\end{equation}
where the bin width $\Delta x = 2^{1-l_\text{GBI}}$ matches the resolution convention used in the BT method (cf. Eq.~\eqref{eq:delta:x}).

\paragraph{Iteration scheme.}
Starting from a uniform distribution, we iterate the discrete transfer operator:
\begin{equation}
P^{(n+1)}_j = \frac{1}{2} \sum_{k: j_k^+ = j} P^{(n)}_k + \frac{1}{2} \sum_{k: j_k^- = j} P^{(n)}_k,
\end{equation}
where $n$ labels the iteration and $j_k^\pm$ are the target bin indices. 
The iteration continues until the $L_1$ convergence criterion
\begin{equation}
  \sum_j |P^{(n+1)}_j - P^{(n)}_j| < \epsilon
\end{equation}
is satisfied, with $\epsilon= 10^{-10}$, or until a maximum of $N_{\max} = 5000$ iterations is reached. After convergence, the distribution is normalized and the entropy is computed
using Eq.~\eqref{eq:entropy:delta:x}. \\
\bibliography{z_all}

@article{Arif2025,
  title = {Impact of Particle-Size Polydispersity on the Quality of Thin-Film Colloidal Crystals},
  author = {Arif, Mariam and Schofield, Andrew B. and Laidlaw, Fraser H. J. and Poon, Wilson C. K. and Thijssen, Job H. J.},
  year = 2025,
  month = oct,
  journal = {Soft Matter},
  volume = {21},
  number = {42},
  pages = {8122--8129},
  publisher = {The Royal Society of Chemistry},
  issn = {1744-6848},
  doi = {10.1039/D5SM00754B},
  urldate = {2025-12-23},
  langid = {english}
}

@article{Corominas-Murtra2019,
  title = {Thermodynamics of Duplication Thresholds in Synthetic Protocell Systems},
  author = {Corominas-Murtra, Bernat},
  year = 2019,
  journal = {Life},
  volume = {9},
  number = {9},
  doi = {10.3390/life9010009}
}

@article{Huang2017,
  title = {Formation and Size Distribution of Self-Assembled Vesicles},
  author = {Huang, Changjin and Quinn, David and Sadovsky, Yoel and Suresh, Subra and Hsia, K. Jimmy},
  year = 2017,
  journal = {Proceedings of the National Academy of Sciences of the USA},
  volume = {114},
  number = {11},
  pages = {2910--2915},
  doi = {10.1073/pnas.1702065114}
}

@article{Krapivsky2004,
  title = {Random Walk with Shrinking Steps},
  author = {Krapivsky, P. L. and Redner, S.},
  year = 2004,
  journal = {American Journal of Physics},
  volume = {72},
  number = {5},
  pages = {591--598},
  doi = {10.1119/1.1632487}
}

@article{Metzler2000,
  title = {The Random Walk's Guide to Anomalous Diffusion: A Fractional Dynamics Approach},
  shorttitle = {The Random Walk's Guide to Anomalous Diffusion},
  author = {Metzler, Ralf and Klafter, Joseph},
  year = 2000,
  month = dec,
  journal = {Physics Reports},
  volume = {339},
  number = {1},
  pages = {1--77},
  issn = {0370-1573},
  doi = {10.1016/S0370-1573(00)00070-3},
  urldate = {2025-12-23}
}

@article{Ozawa2017,
  title = {Does the Configurational Entropy of Polydisperse Particles Exist?},
  author = {Ozawa, Misaki and Berthier, Ludovic},
  year = 2017,
  journal = {The Journal of Chemical Physics},
  volume = {146},
  number = {1},
  pages = {014502},
  doi = {10.1063/1.4972525}
}

@article{Pusey1987,
  title = {The Effect of Polydispersity on the Crystallization of Hard Spherical Colloids},
  author = {Pusey, P. N.},
  year = 1987,
  journal = {Journal de Physique (France)},
  volume = {48},
  number = {5},
  pages = {709--712},
  doi = {10.1051/jphys:01987004805070900}
}

@article{Schrum2010,
  title = {The Origins of Cellular Life},
  author = {Schrum, Jason P. and Zhu, Ting F. and Szostak, Jack W.},
  year = 2010,
  month = sep,
  journal = {Cold Spring Harbor Perspectives in Biology},
  volume = {2},
  number = {9},
  pages = {a002212},
  issn = {1943-0264},
  doi = {10.1101/cshperspect.a002212},
  urldate = {2025-12-15},
  pmcid = {PMC2926753},
  pmid = {20484387}
}

@misc{Varju2016,
      title={Recent progress on {B}ernoulli convolutions}, 
      author={Peter P. Varju},
      year={2016},
      eprint={1608.04210},
      archivePrefix={arXiv},
      primaryClass={math.CA},
      url={https://arxiv.org/abs/1608.04210}
}

@misc{Shivers2025,
  title = {Renormalized Mechanics and Stochastic Thermodynamics of Growing Model Protocells},
  author = {Shivers, Jordan L. and Nguyen, Michael and Dinner, Aaron R. and Vlahovska, Petia and Vaikuntanathan, Suriyanarayanan},
  year = 2025,
  eprint = {2503.24120},
  primaryclass = {cond-mat.soft},
  doi = {10.48550/arXiv.2503.24120},
  archiveprefix = {arXiv},
  howpublished = {arXiv preprint}
}

@article{Sollich2002,
  title = {Predicting Phase Equilibria in Polydisperse Systems},
  author = {Sollich, Peter},
  year = 2002,
  month = jan,
  journal = {Journal of Physics: Condensed Matter},
  volume = {14},
  number = {3},
  pages = {R79--R117},
  issn = {0953-8984, 1361-648X},
  doi = {10.1088/0953-8984/14/3/201},
  urldate = {2025-12-19},
  langid = {english}
}

@book{Weiss1994,
  title = {Aspects and Applications of the Random Walk},
  author = {Weiss, George H.},
  year = 1994,
  series = {Random Materials and Processes},
  publisher = {North-Holland},
  address = {Amsterdam},
  isbn = {978-0-444-81606-1}
}

@article{Zwicker2017,
  title = {Growth and Division of Active Droplets Provides a Model for Protocells},
  author = {Zwicker, David and Seyboldt, Rabea and Weber, Christoph A. and Hyman, Anthony A. and J{\"u}licher, Frank},
  year = 2017,
  journal = {Nature Physics},
  volume = {13},
  pages = {408--413},
  doi = {10.1038/nphys3984}
}

@article{Alexander1991,
  title = {The Entropy of a Certain Infinitely Convolved {B}ernoulli Measure},
  author = {Alexander, J. C. and Zagier, Don},
  year = 1991,
  month = aug,
  journal = {Journal of the London Mathematical Society},
  volume = {s2-44},
  number = {1},
  pages = {121--134},
  issn = {00246107},
  doi = {10.1112/jlms/s2-44.1.121},
  urldate = {2025-08-09},
  copyright = {http://doi.wiley.com/10.1002/tdm\_license\_1.1},
  langid = {english}
}

@article{Amir2014,
  title = {Cell Size Regulation in Bacteria},
  author = {Amir, Ariel},
  year = 2014,
  month = may,
  journal = {Physical Review Letters},
  volume = {112},
  number = {20},
  pages = {208102},
  publisher = {American Physical Society},
  doi = {10.1103/PhysRevLett.112.208102},
  urldate = {2025-03-05}
}

@article{Barkai1999b,
  title = {Distribution of Single-Molecule Line Widths},
  author = {Barkai, E and Silbey, R},
  year = 1999,
  month = sep,
  journal = {Chemical Physics Letters},
  volume = {310},
  number = {3},
  pages = {287--295},
  issn = {0009-2614},
  doi = {10.1016/S0009-2614(99)00797-6},
  urldate = {2026-01-08}
}

@article{Ben-Naim1992,
  title = {Bimodal Diffusion in Power-Law Shear Flows},
  author = {{Ben-Naim}, E. and Redner, S. and {ben-Avraham}, D.},
  year = 1992,
  month = may,
  journal = {Physical Review A},
  volume = {45},
  number = {10},
  pages = {7207--7213},
  publisher = {American Physical Society},
  doi = {10.1103/PhysRevA.45.7207},
  urldate = {2026-01-08}
}

@article{Elgamel2024,
  title = {Effects of Molecular Noise on Cell Size Control},
  author = {Elgamel, Mohamed and Mugler, Andrew},
  year = 2024,
  journal = {Physical Review Letters},
  volume = {132},
  number = {9},
  pages = {098403},
  publisher = {American Physical Society},
  doi = {10.1103/PhysRevLett.132.098403}
}

@article{Grassberger1983,
  title = {Characterization of Strange Attractors},
  author = {Grassberger, Peter and Procaccia, Itamar},
  year = 1983,
  month = jan,
  journal = {Physical Review Letters},
  volume = {50},
  number = {5},
  pages = {346--349},
  issn = {0031-9007},
  doi = {10.1103/PhysRevLett.50.346},
  urldate = {2026-01-15},
  copyright = {http://link.aps.org/licenses/aps-default-license},
  langid = {english}
}

@article{KozachenkoLeonenko1987,
  title = {Sample Estimate of the Entropy of a Random Vector},
  author = {Kozachenko, L. F. and Leonenko, N. N.},
  year = 1987,
  journal = {Problems of Information Transmission},
  volume = {23},
  number = {2},
  pages = {95--101}
}

@book{LandauLifshitz1980StatPhys1,
  title = {Statistical Physics, Part 1},
  author = {Landau, L. D. and Lifshitz, E. M.},
  year = {1980},
  series = {Course of Theoretical Physics},
  edition = {3},
  volume = {5},
  publisher = {Elsevier, Oxford, UK},
  isbn = {978-0-08-023039-9}
}

@book{Chandler1987,
  title = {Introduction to Modern Statistical Mechanics},
  author = {David Chandler},
  year = {1987},
  publisher = {Oxford University Press, Oxford, UK}
}

@book{Yeomans1992,
  title = {Statistical Mechanics of Phase Transitions},
  author = {J. M. Yeomans},
  year = {1992},
  publisher = {Oxford University Press, Oxford, UK}
}

@book{Klafter2011,
  title = {First Steps in Random Walks: {F}rom Tools to Applications},
  author = {J. Klafter and I.M. Sokolov},
  year = {2011},
  publisher = {Oxford University Press, Oxford, UK}
}

@incollection{Peres2000,
  title = {Sixty Years of {B}ernoulli Convolutions},
  booktitle = {Fractal Geometry and Stochastics II},
  author = {Peres, Yuval and Schlag, Wilhelm and Solomyak, Boris},
  editor = {Bandt, Christoph and Graf, Siegfried and Z{\"a}hle, Martina},
  year = 2000,
  pages = {39--65},
  publisher = {Birkh\"auser Basel},
  address = {Basel},
  doi = {10.1007/978-3-0348-8380-1_2},
  urldate = {2025-08-08},
  isbn = {978-3-0348-9542-2 978-3-0348-8380-1},
  langid = {english}
}

@article{Pikovsky2023,
  title = {Statistical Theory of Asymmetric Damage Segregation in Clonal Cell Populations},
  author = {Pikovsky, Arkady and Tsimring, Lev S.},
  year = 2023,
  month = apr,
  journal = {Mathematical Biosciences},
  volume = {358},
  pages = {108980},
  issn = {0025-5564},
  doi = {10.1016/j.mbs.2023.108980},
  urldate = {2026-01-08}
}

@article{Sornette1998,
  title = {Discrete-Scale Invariance and Complex Dimensions},
  author = {Sornette, D.},
  year = 1998,
  journal = {Physics Reports},
  volume = {297},
  number = {5},
  pages = {239--270},
  doi = {10.1016/S0370-1573(97)00076-8}
}

@book{Bishop_2006,
title={Pattern Recognition and Machine Learning},
publisher={Springer Science+Business Media, LLC, New York, NY, USA},
author={Bishop, Christopher M.},
year={2006}
}

@article{Chen2006TimeSpaceFabric,
  author  = {Chen, W.},
  title   = {Time-space fabric underlying anomalous diffusion},
  journal = {Chaos, Solitons \& Fractals},
  year    = {2006},
  volume  = {28},
  number  = {4},
  pages   = {923--929},
  doi     = {10.1016/j.chaos.2005.08.199},
  issn    = {0960-0779}
}

@article{Erdos1939,
  title = {On a Family of Symmetric {B}ernoulli Convolutions},
  author = {Erd{\H o}s, P.},
  year = 1939,
  journal = {American Journal of Mathematics},
  volume = {61},
  number = {4},
  pages = {974--974},
  doi = {10.2307/2371641}
}

\end{document}